\documentclass[prl,twocolumn,superscriptaddress,floatfix]{revtex4-1}
\usepackage{graphicx}
\usepackage{color}
\usepackage{dcolumn}
\usepackage{bm}
\usepackage{amssymb}
\usepackage{color,amsmath}
\usepackage{soul,xcolor}
\setstcolor{red}
\usepackage{epstopdf}

\newcommand{\tf}{\mathrm{TF}}

\newcommand{\nmsq}{\mathrm{nm}^2}

\begin{document}
\title{Imaging orbital ferromagnetism in a moir\'e Chern insulator}
\author{C. L. Tschirhart}
\thanks{These two authors contributed equally}
\affiliation{Department of Physics, University of California, Santa Barbara, CA 93106, USA}
\author{M. Serlin}
\thanks{These two authors contributed equally}
\affiliation{Department of Physics, University of California, Santa Barbara, CA 93106, USA}
\author{H. Polshyn}
\affiliation{Department of Physics, University of California, Santa Barbara, CA 93106, USA}
\author{A. Shragai}
\thanks{Current address: Department of Physics, Cornell University, Ithaca, NY 14853, USA}
\affiliation{Department of Physics, University of California, Santa Barbara, CA 93106, USA}
\author{Z. Xia}
\affiliation{Department of Physics, University of California, Santa Barbara, CA 93106, USA}
\author{J. Zhu}
\thanks{Current address: Department of Physics, Cornell University, Ithaca, NY 14853, USA}
\affiliation{Department of Physics, University of California, Santa Barbara, CA 93106, USA}
\author{Y. Zhang}
\affiliation{Department of Physics, University of California, Santa Barbara, CA 93106, USA}
\author{K. Watanabe}
\affiliation{Research Center for Functional Materials, National Institute for Materials Science, 1-1 Namiki, Tsukuba 305-0044, Japan}
\author{T. Taniguchi}
\affiliation{International Center for Materials Nanoarchitectonics, National Institute for Materials Science, 1-1 Namiki, Tsukuba 305-0044, Japan}
\author{M. E. Huber}
\affiliation{Departments of Physics and Electrical Engineering, University of Colorado Denver, Denver, CO
80217, USA}
\author{A. F. Young}
\email{andrea@physics.ucsb.edu}
\affiliation{Department of Physics, University of California, Santa Barbara, CA 93106, USA}

\begin{abstract}
Electrons in moir\'e flat band systems can spontaneously break time reversal symmetry, giving rise to a quantized anomalous Hall effect. Here we use a superconducting quantum interference device to image stray magnetic fields in one such system composed of twisted bilayer graphene aligned to hexagonal boron nitride. We find a magnetization of several Bohr magnetons per charge carrier, demonstrating that the magnetism is primarily orbital in nature.  Our measurements reveal a large change in the magnetization as the chemical potential is swept across the quantum anomalous Hall gap consistent with the expected contribution of chiral edge states to the magnetization of an orbital Chern insulator. Mapping the spatial evolution of field-driven magnetic reversal, we find a series of reproducible micron scale domains whose boundaries host chiral edge states.

\end{abstract}
\maketitle
In crystalline solids, Berry curvature endows the energy bands with an orbital magnetic moment\cite{xiao_berry_2010}. While the orbital moment often contributes---at times substantially\cite{adachi_ferromagnet_1999}---to the net magnetization of ferromagnets, all known ferromagnetic materials involve partial or full polarization of the electron spin.
Theoretically, however, ferromagnetism can also arise through the spontaneous polarization of orbital magnetic moments without involvement of the electron spin.
Recently, hysteretic transport consistent with ferromagnetic order has been observed in  heterostructures composed of graphene and hexagonal boron nitride\cite{sharpe_emergent_2019,serlin_intrinsic_2020,chen_tunable_2019,polshyn_nonvolatile_2020,chen_electrically_2020}, neither of which are intrinsically magnetic materials.  Notably,   spin-orbit coupling is thought to be vanishingly small in these systems\cite{sichau_resonance_2019}, effectively precluding a spin-based mechanism. These results have consequently been interpreted as evidence for the first purely orbital ferromagnet\cite{xie_nature_2020,bultinck_anomalous_2019,zhang_twisted_2019,liu_correlated_2020,wu_collective_2020,chatterjee_symmetry_2019,repellin_ferromagnetism_2019,alavirad_ferromagnetism_2019}.

To host purely orbital ferromagnetic order, a system must have a time reversal symmetric electronic degree of freedom separate from the electron spin as well as strong electron-electron interactions. Both are present in graphene heterostructures, where the valley degree of freedom provides degenerate electron species related by time reversal symmetry and a moir\'e superlattice can be used to engineer strong interactions. In these materials, a long wavelength moir\'e pattern, arising from interlayer coupling between mismatched lattices, modulates the underlying electronic structure and leads to the emergence of superlattice minibands within a reduced Brillouin zone. The small Brillouin zone means that only low electron densities are required to dope the 2D system to full filling or depletion of the superlattice bands, which can be achieved using experimentally realizable electric fields\cite{kim_tunable_2017}.  For appropriately chosen constituent materials and interlayer rotational alignment, the lowest energy bands can have bandwidths considerably smaller than the native scale of electron-electron interactions, $E_C\approx e^2/\lambda_M$, where $\lambda_M$ is the moir\'e period.
The dominance of interactions typically manifests experimentally through the appearance of `correlated insulators' at integer electron or hole filling of the moir\'e unit cell\cite{cao_correlated_2018,chen_evidence_2019}, consistent with interaction-induced breaking of one or more of the  spin, valley, or lattice symmetries.  Orbital magnets are thought to constitute a subset of these states, in which exchange interactions favor a particular order that breaks time-reversal symmetry by causing the system to polarize into one or more valley projected bands.  Remarkably, the large Berry curvature endows the valley projected bands with a finite Chern number\cite{song_topological_2015,zhang_nearly_2019}, so that valley polarization naturally leads to a quantized anomalous Hall effect at integer band filling. To date, quantum anomalous Hall effects have been observed at band fillings $\nu=1$ and $\nu=3$ in various heterostructures\cite{serlin_intrinsic_2020,polshyn_nonvolatile_2020,chen_tunable_2020}, where $\nu=A n$ corresponds to the number of electrons per unit cell area $A$ with $n$ the carrier density.

\begin{figure*}[h]
\includegraphics[width=7.25in]{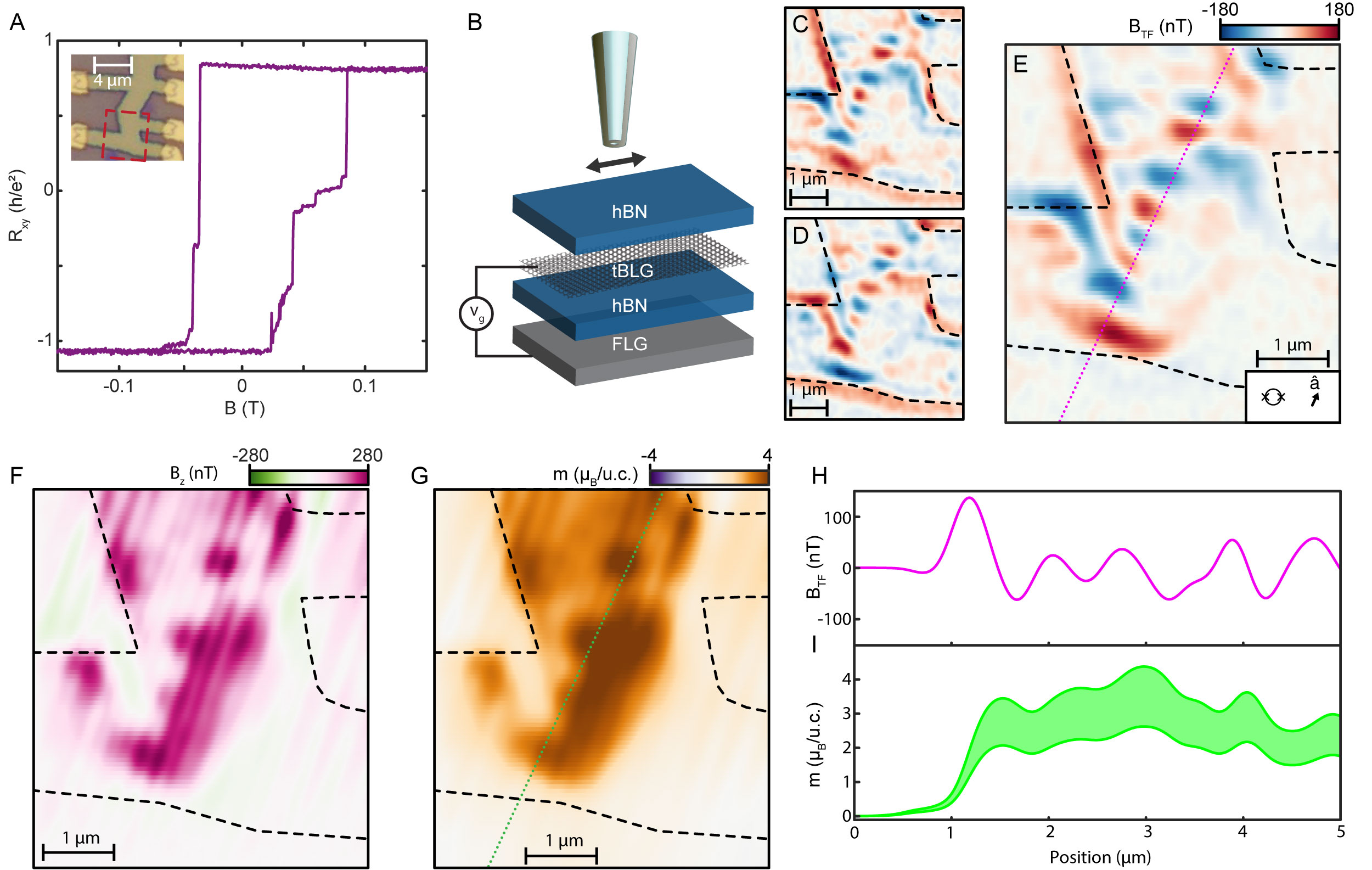}
\caption{
\textbf{Imaging orbital ferromagnetism.}
(\textbf{A}) Hall resistance of a twisted bilayer graphene device aligned to hexagonal boron nitride measured at T = 2.1~K and electron density $n = 2.36 \times 10^{12}$ cm$^{-2}$. A thorough characterization of transport in this exact device is reported in Ref. \onlinecite{serlin_intrinsic_2020}.
Inset: optical micrograph of the device with the scan region marked in red. The scale bar is $4~\mu$m.
(\textbf{B})  Schematic illustration of the experimental setup.
We raster a nanoscale indium SQUID with diameter $d = 215$~nm at a height of $h\approx 140$~nm above the plane of the twisted bilayer graphene.
The SQUID is coupled to a quartz tuning fork whose excitation causes the tip to oscillate mechanically at $f_\tf\approx 33$~kHz in the plane of the sample. The SQUID response at this frequency $B_\tf\approx \hat a\cdot \vec \nabla_r B_{z}$ where $B_z$ is the static magnetic field and $|\hat a|\approx 189$~nm is the tuning fork oscillation amplitude.
(\textbf{C})
$B_\tf$ measured at B=22~mT after field training to +200~mT and
(\textbf{D}) -200~mT.
(\textbf{E}) Half the difference between data shown in panels C and D. Panels C, D and E share the same color scale. Inset: to-scale representations of the SQUID diameter and tuning fork amplitude $\hat a$.
(\textbf{F}) $B_z$ as determined by integrating data in panel E. We assume $B_z=0$ along the bottom and left edges of the scan range.
(\textbf{G})
Magnetization density $m$.  Data are presented in units of Bohr magnetons per moir\'e unit cell area $A\approx$ 130 nm$^2$\cite{serlin_intrinsic_2020}.
(\textbf{H})
$B_\tf$ and
(\textbf{I}) $m$ plotted along the indicated contours in panels E and G.  The shaded regions in H show bounds on $m$ obtained by propagating uncertainty in $|\hat a|$ (see SI). }
\label{fig:imaging}
\end{figure*}

While orbital magnetism is generally expected theoretically, no direct experimental probes of magnetism have been reported due to the relative scarcity of magnetic samples, their small size, and the low expected magnetization density of one orbital moment per superlattice cell.  The resulting magnetization density $m\lesssim 0.1  ~\mu_B/\nmsq$\cite{zhu_curious_2020} (where $\mu_B\approx 0.06$ meV/T is
the Bohr magneton) is consequently over three orders of magnitude smaller than in typical magnetic systems with several spins per subnanometer-sized crystal unit cell.
The absence of magnetic studies leaves open both quantitative questions, such as the size of the orbital moment, as well as qualitative ones regarding the nature of the magnetic phase transitions as a function of magnetic field and carrier density.

Here, we perform spatially resolved magnetometry to image the submicron magnetic structure of the same sample presented in Reference \onlinecite{serlin_intrinsic_2020} (see Fig.~\ref{fig:imaging}A), which consists of a twisted graphene bilayer aligned to one of the hexagonal boron nitride encapsulating layers.
Figure \ref{fig:imaging}B shows a schematic representation of our experimental setup.  We use a superconducting quantum interference device (SQUID) fabricated on the tip of a quartz tube from cryogenically deposited indium\cite{anahory_squid--tip_2020} with a magnetic field sensitivity of approximately $15$ nT/Hz$^{1/2}$ at select out-of-plane magnetic fields of less than 50 mT (see Fig. \ref{fig:nanosquid}).
The SQUID is mounted to a quartz tuning fork \cite{uri_nanoscale_2020} (see Fig. \ref{fig:tf_intro}) and rastered in a 2D plane parallel, and at a fixed height above, the tBLG heterostructure.
A finite electrical excitation applied to the tuning fork generates a lateral oscillation of the tip along vector $\hat a$, and we measure the SQUID response at the tuning fork oscillation frequency, $B_\tf\approx \hat a\cdot \vec \nabla_r B_{z}$ (see Fig. \ref{fig:tf_cal}).

Figs. \ref{fig:imaging}C and D show images of $B_\tf$ taken while the sample is doped to $n = 2.36 \times 10^{12}$ cm$^{-2}$, near the quantized Hall plateau corresponding to $\nu=3$.
Images are acquired in the same background magnetic field $B=22$~mT but on opposite branches of the hysteresis loop shown in Fig. \ref{fig:imaging}A.
As discussed in the SI and Fig. \ref{fig:parasitics}, the measured $B_{TF}$ contains contributions from both magnetic signals as well as other effects arising from electric fields or thermal gradients. To isolate the magnetic structure that gives rise to the observed hysteretic transport, we subtract data from Figs. \ref{fig:imaging}C and D from each other.  The result is shown in Fig.~\ref{fig:imaging}E, which depicts the gradient magnetometry signal associated with the fully polarized orbital ferromagnet.
To reconstruct the static out of plane magnetic field, $B_z$, we then integrate $B_\tf$ along $\hat{a}$ from the lower and left boundaries of the image (Fig.~\ref{fig:imaging}F).
We infer the total magnetization density $m$ from the $B_z$ data using standard Fourier domain techniques (see SI) as shown in Fig.~\ref{fig:imaging}G.
Figures~\ref{fig:imaging}H-I show a comparison of $B_\tf$ and $m$ plotted along the contours indicated in Figs. \ref{fig:imaging}E and G.  The shaded regions in Fig. \ref{fig:imaging}I denote confidence intervals, whose size is dominated by systematic uncertainty in $|\hat a|$.

Our measurements are taken close to $\nu=3$, equivalent to a single hole per unit cell relative to the nonmagnetic state at $\nu=4$ that corresponds to full filling of the lowest energy bands.
We find that the magnetization density is considerably larger than $1~ \mu_B$ per unit cell area $A\approx 130$~$\nmsq$.
Without any assumptions about the nature of the broken symmetries, this state has a maximum spin magnetization of 1 $\mu_{B}$ per moir\'e unit cell.
Our data reject this hypothesis, finding instead a maximum magnetization density of $m$ in the range  $2-4$ $\mu_{B}$ per moir\'e unit cell corresponding to an orbital moment of 1.8-3.6$\times$ 10$^{-4} \mu_B/$carbon atom.
We conclude that the magnetic moment associated with the QAH phase in tBLG is dominated by its orbital component.

\begin{figure}[ht!]
\includegraphics[width=2.25in]{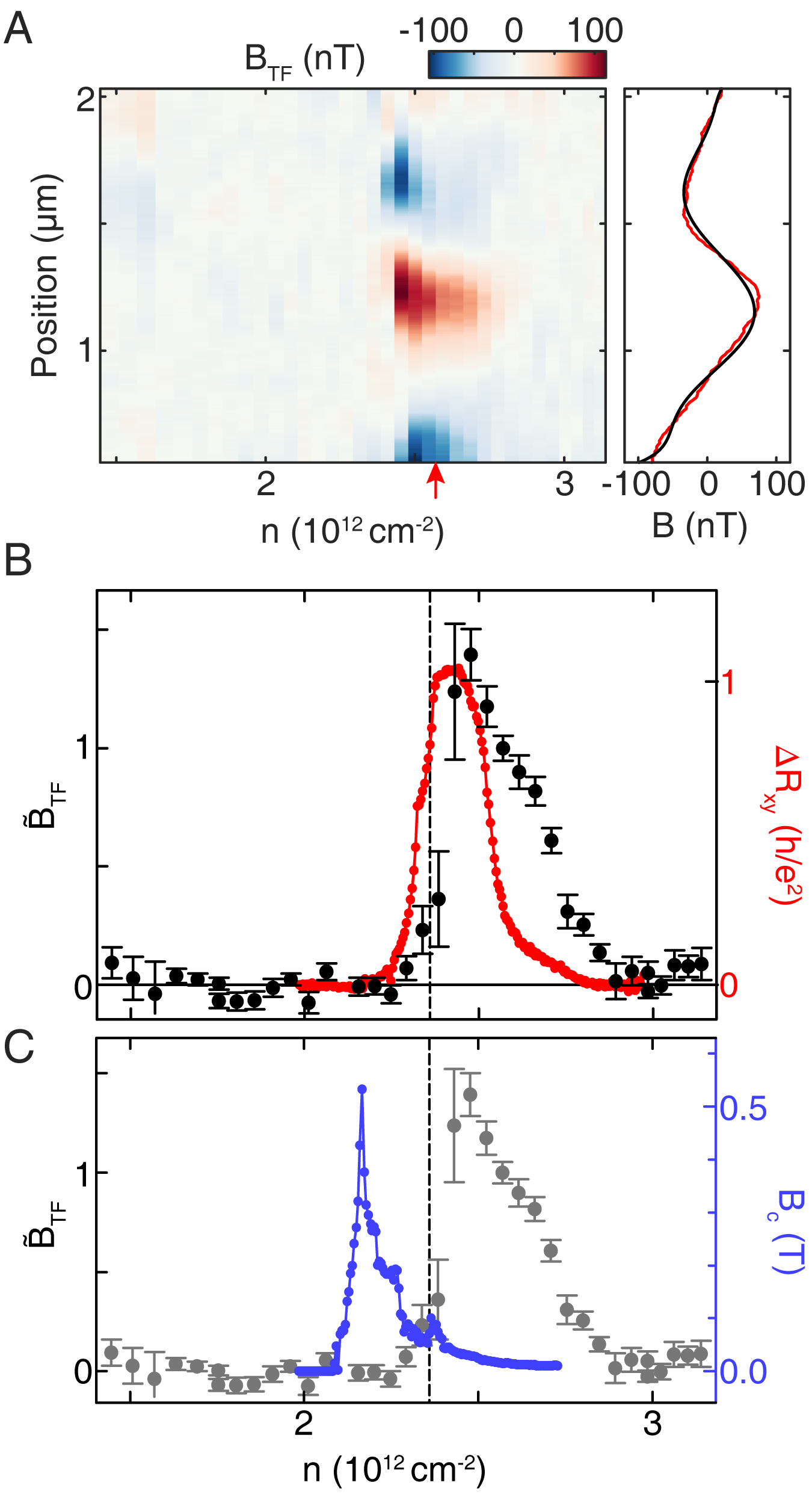}
\caption{\textbf{Density dependence of magnetization.}
(\textbf{A})
Evolution of $B_\tf$ with $n$ in the vicinity of $\nu=3$, measured along the contour shown in Fig.~\ref{fig:imaging}E.  The data was taken at B = 44~mT and T = 2.2~K.
The trace corresponding to $n$ = 2.57$\mathrm{\times 10^{12}cm^{-2}}$ is shown in red at right, along with a fit to a 7th order polynomial in black.
(\textbf{B}) Comparison of magnetic signal with the residual Hall resistance $\Delta R_{xy}$, shown in red.
$B_\tf$ traces at different $n$ are fit to the same polynomial as in panel (A) with a scale factor $\tilde B_\tf$, which serves as a proxy for $m$. Error bars measure standard error of the mean of the residuals of these fits.
(\textbf{C})
Coercive field $B_c$ determined from transport measurements plotted alongside $\tilde B_\tf$.
}
\label{fig:density}
\end{figure}

\begin{figure*}[t!]
\includegraphics[width=4.75in]{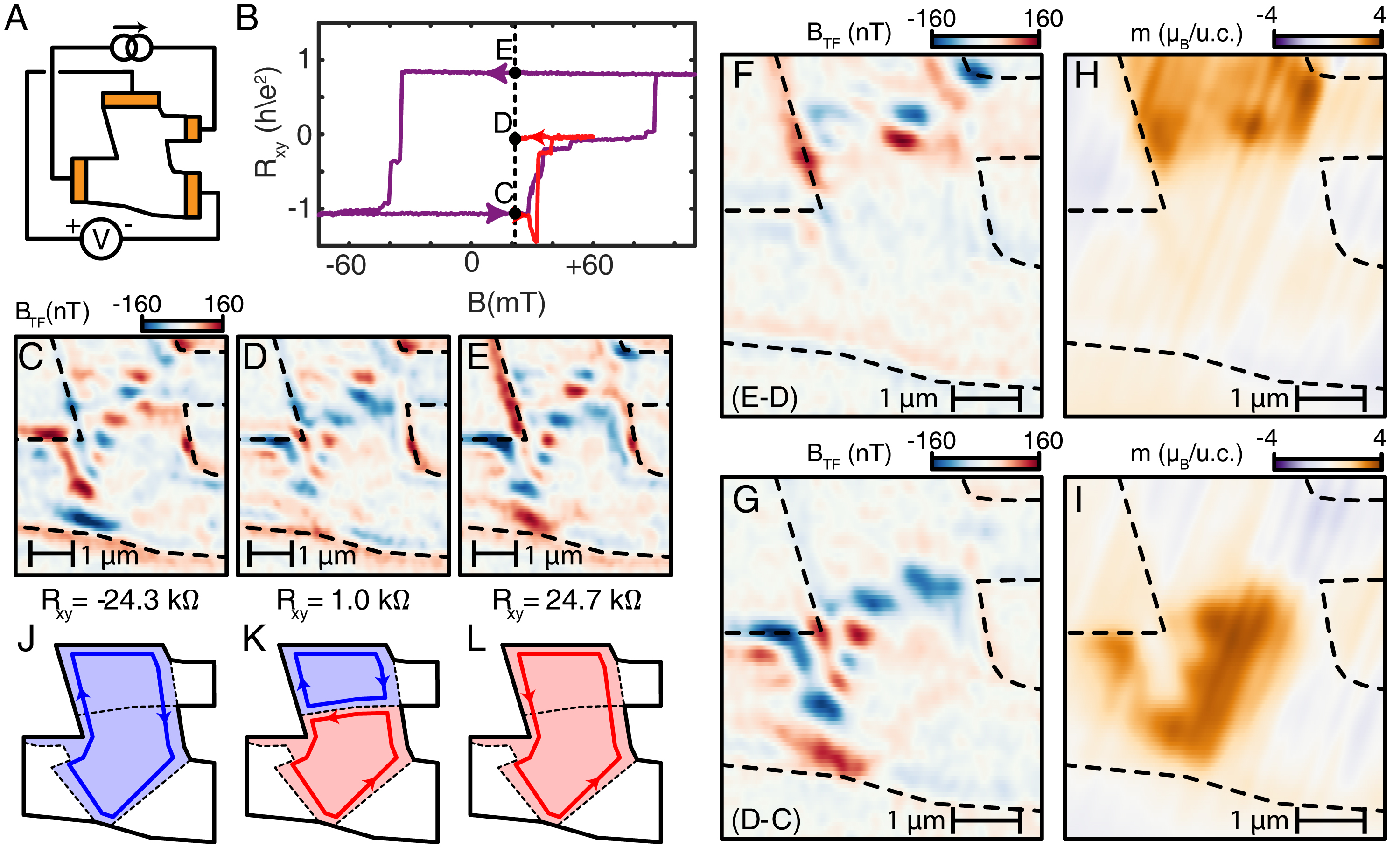}
\caption{
\textbf{Mesoscopic magnetic domains.}
(\textbf{A}) Transport measurement schematic and
(\textbf{B}) Hall resistance data as a function of magnetic field for major (purple) and minor (red) hysteresis loops. (\textbf{C}-\textbf{E}) Gradient magnetometry images of the tBLG device at the three indicated values of the Hall resistance. All magnetic imaging presented here was performed at $B = 22$~mT, $T = 2.1$~K, and $n = 2.36 \times 10^{12}$ cm$^{-2}$.
(\textbf{F}-\textbf{G}) Pairwise difference images based on the data presented in panels (C-E). The same domain structures were observed on multiple cooldowns, reminiscent of grains formed by crystalline domain walls in polycrystalline ferromagnetic metals.
(\textbf{H}-\textbf{I}) Extracted magnetization $m$ for the images in panels (F-G).
(\textbf{J}-\textbf{L}) Schematic depiction of edge state structure corresponding to magnetization states in C-E. Assuming full edge state equilibration, these states would result in $R_{xy}=-h/e^2$, 0, and $h/e^2$, close to the observed values. }
\label{fig:grains}
\end{figure*}

The net orbital magnetic moment of a set of Bloch wave functions is a function of the Berry curvature of the filled states.  Because this depends strongly on band filling, the orbital magnetic moment can vary dramatically in response to changes in electron density. To explore the density dependence of the local magnetic structure, Figure \ref{fig:density}A shows repeated measurements of $B_\tf$ for a series of gate voltages in the vicinity of $\nu=3$. $B_\tf$ is measured along a contour that runs over a region of the device showing magnetic inhomogeneity even at the saturation magnetization (see Figs. \ref{fig:imaging}E and H). Assuming that the magnetic structure is density independent, the amplitude of the position dependent modulation of $B_\tf$ functions as a proxy for $m$.

To compare magnetization at different $n$, we fit the data for $n$ = 2.57$\mathrm{\times 10^{12}cm^{-2}}$ to a 7th order polynomial (see Fig. \ref{fig:density}A); all other curves are then fit to the same polynomial with an overall scale factor which we denote $\tilde B_\tf$ and plot in Fig. \ref{fig:density}B.  Error bars reflect standard error of the mean for these fits.  Comparing the $n$-dependent  $\tilde B_\tf$ with the residual Hall resistance shows that $\tilde B_\tf$ grows slowly as density is lowered towards $\nu=3$, but then abruptly drops below the noise floor of our SQUID measurements within the range of $n$ associated with the quantized $R_{xy}$ plateau.

The dramatic change of the inferred $m$ within the quantum anomalous Hall plateau is in line with expectations for orbital Chern insulators.
Due to the chiral edge states, Chern insulators feature a chemical-potential dependent contribution to the magnetization even within the bulk energy gap corresponding to the quantized transport plateau\cite{zhu_curious_2020}. This leads to a jump in magnetization across a Chern insulator gap of $\Delta m=C E_{gap}/\Phi_0$, where $C$ is the Chern number and  $\Phi_0=h/e$ is the flux quantum. Transport measurements of the quantum anomalous Hall state at $\nu=3$ in this sample found $E_{gap}= 2.5$ meV\cite{serlin_intrinsic_2020}, corresponding to a $\Delta m$ of $1.4~\mu_B$ per unit cell. The quantitative discrepancy between $\delta m\gtrsim 3$ inferred from the collapse of $\tilde B_\tf$ near the gap may arise from disorder effects, which typically lead transport experiments to underestimate $E_g$.  Alternatively, if the internal magnetic structure of the device is strongly $n$-dependent, the linecut measurement $\tilde B_\tf$ may not be a good proxy for the maximum value of $m$.  Nevertheless, the qualitative behavior of the magnetic signal in the quantum anomalous Hall plateau is strikingly similar to that predicted for orbital Chern insulators, and strikingly different from observations in spin Chern insulators where $m$ is $n$-independent\cite{lachman_visualization_2015}.

Though topological edge states contribute to the magnetization of all Chern insulators, the low bulk moment density of orbital magnets allows this contribution to be comparable to that of the bulk, in some cases leading to a vanishing total $m$\cite{zhu_curious_2020,polshyn_quantitative_2018}.
Fig. \ref{fig:density}C compares $\tilde B_\tf$ to the transport coercive field $B_c$. $B_c$ increases rapidly as density is decreased below the transport plateau, reaching a sharp maximum near $n=2.2 \times 10^{12} \mathrm{cm}^{-2}$ , concomitant with a change in sign of the anomalous Hall effect (Fig. \ref{fig:S:signchange}).  The  sign change occurs where our local magnetometry shows  $m$ to be very small. These facts can be reconciled by noting that the coercive field is expected to increase as $m$ decreases, since $m$ sets the coupling to the external magnetic field in the free energy, $m=-\partial F/\partial B$.  Within this picture, the peak in the coercive field corresponds to a sign change in the orbital moment within a given valley, brought about in part by the large contributions of the topological edge states.

Previous work on graphene-based Chern insulators has found Barkhausen noise jumps comparable to $h/e^2$ \cite{serlin_intrinsic_2020, sharpe_emergent_2019,polshyn_nonvolatile_2020}, suggesting a substructure of only a handful of ferromagnetic domains comparable in size to the distance between contacts.  However, our magnetometry data show significant submicron scale inhomogeneity even at full magnetic saturation. This is similar to findings in transition metal doped topological insulators, where the magnetic structure is dominated by inhomogenous distribution and clustering of the Cr or V dopants.  In those systems, magnetic imaging shows superparamagnetic dynamics characterized by the reversal of weakly correlated point-like microscopic magnetic dipoles\cite{lachman_visualization_2015,lachman_observation_2017,yasuda_quantized_2017}.
Transport, meanwhile, does not typically show substantial Barkhausen noise\cite{chang_experimental_2013}, with the exception of one report where jumps were reported in a narrow range of temperatures\cite{liu_large_2016}.

To investigate the domain dynamics directly, we compare magnetic structure across different states stabilized in the midst of magnetic field driven reversal.
Fig.~\ref{fig:grains}A shows a schematic depiction of our transport measurement, and Fig. \ref{fig:grains}B shows the resulting$R_{xy}$ data for both a major hysteresis loop spanning the two fully polarized states at $R_{xy}=\pm h/e^2$ (in purple) and a minor loop that terminates in a mixed polarization state at $R_{xy}\approx 0$ (in red).
All three states represented by these hysteresis loops can be stabilized at $B=22$~mT for T = 2.1~K, where our nanoSQUID has excellent sensitivity, allowing a direct comparison of their respective magnetic structures (Figs.~\ref{fig:grains}C-E).
Figs. \ref{fig:grains}F-G show images obtained by subtracting one of the images at full positive or negative polarization from the mixed state.
Applying the same magnetic inversion algorithm used in Fig. \ref{fig:imaging} produces maps of $m$ corresponding to these differences (Figs. \ref{fig:grains}H-I), allowing us to visualize the domain structure generating the intermediate plateau $R_{xy}\approx 0$ seen in the major hysteresis loop. Evidently, the Hall resistance of the device in this state is dominated by the interplay of two large magnetic domains, each comprising about half of the active area.

Armed with knowledge of the domain structure, it is straightforward to understand the behavior of the measured transport in the mixed state imaged in Fig. \ref{fig:grains}D. In particular, the state corresponds to the presence of a single domain wall that crosses the device, separating both the current and the Hall voltage contacts (see Fig. \ref{fig:grains}A).  In the limit in which the topological edge states at the boundaries of each magnetic domain are in equilibrium, there will be no drop in chemical potential across the domain wall, leading to $R_{xy} = 0$.  This is very close to the observed value of $R_{xy} = 1.0 ~\text{k}\Omega = 0.039~ \text{h}/\text{e}^2$. As shown in Figs. \ref{fig:trans_repeat} and \ref{fig:S:moregrains}, more subtle features of the transport curve can also be associated with the reversal of domains which do not bridge contacts.

We find that the observed domain reversals associated with the Barkhausen jumps are consistent over repeated thermal cycles between cryogenic and room temperature. Together, these findings suggest a close analogy to polycrystalline spin ferromagnets, which host ferromagnetic domain walls that are strongly pinned to crystalline grain boundaries; indeed, these crystalline grains are responsible for Barkhausen noise as it was originally described\cite{barkhausen_zwei_1919}.
While crystalline defects are unlikely in tBLG due to the high degree of perfection of the constituent graphene layers, prior studies of a variety of moir\'e systems\cite{balents_superconductivity_2020, uri_mapping_2019}---including transport characterization of the present device\cite{serlin_intrinsic_2020}---all show significant spatial variations in the moir\'e pattern itself.  Adding additional complication is the possibility of a spatially varying coupling between the tBLG layers and the hBN substrate, which is thought to stabilize magnetism. Despite the current lack of experimental control over moir\'e disorder, the reproducibility of the grain structure suggests that deterministic engineering of coercive fields and magnetization patterns through boundary and defect control will be possible.

\section*{Acknowledgments}
The authors thank A.H. Macdonald, J. Zhu, M. Zaletel, and D. Xiao for discussions of the results and E. Lachman for their comments on the manuscript.
The work was primarily funded by the Department of Energy under DE-SC0020043, with additional support for instrumentation development supported by the Army Research Office under Grant No. W911NF-16-1-0361. KW and TT acknowledge support from the Elemental Strategy Initiative conducted by the MEXT, Japan, Grant Number JPMXP0112101001, JSPS KAKENHI Grant Numbers JP20H00354 and the CREST (JPMJCR15F3), JST.
CLT acknowledges support from the Hertz Foundation and from the National Science Foundation Graduate Research Fellowship Program under grant 1650114.
This project is funded in part by the Gordon and Betty Moore Foundation’s EPiQS Initiative, Grant GBMF9471 to AFY.

\section*{Competing interests}
The authors declare no competing interests.

\section*{Data availability}
The data that support the plots within this paper and other findings of this study are available from the corresponding author upon reasonable request.

\clearpage

\bibliographystyle{custom}
\bibliography{references}
\clearpage

\pagebreak
\widetext
\begin{center}
\textbf{\large Supplementary Information}
\end{center}
\renewcommand{\thefigure}{S\arabic{figure}}
\renewcommand{\thesubsection}{S\arabic{subsection}}
\setcounter{secnumdepth}{2}
\renewcommand{\theequation}{S\arabic{equation}}
\renewcommand{\thetable}{S\arabic{table}}
\setcounter{figure}{0}
\setcounter{equation}{0}
\onecolumngrid

\section{Methods}

\subsection{NanoSQUID on tip}

\begin{figure}[ht]
\includegraphics[width=4.75in]{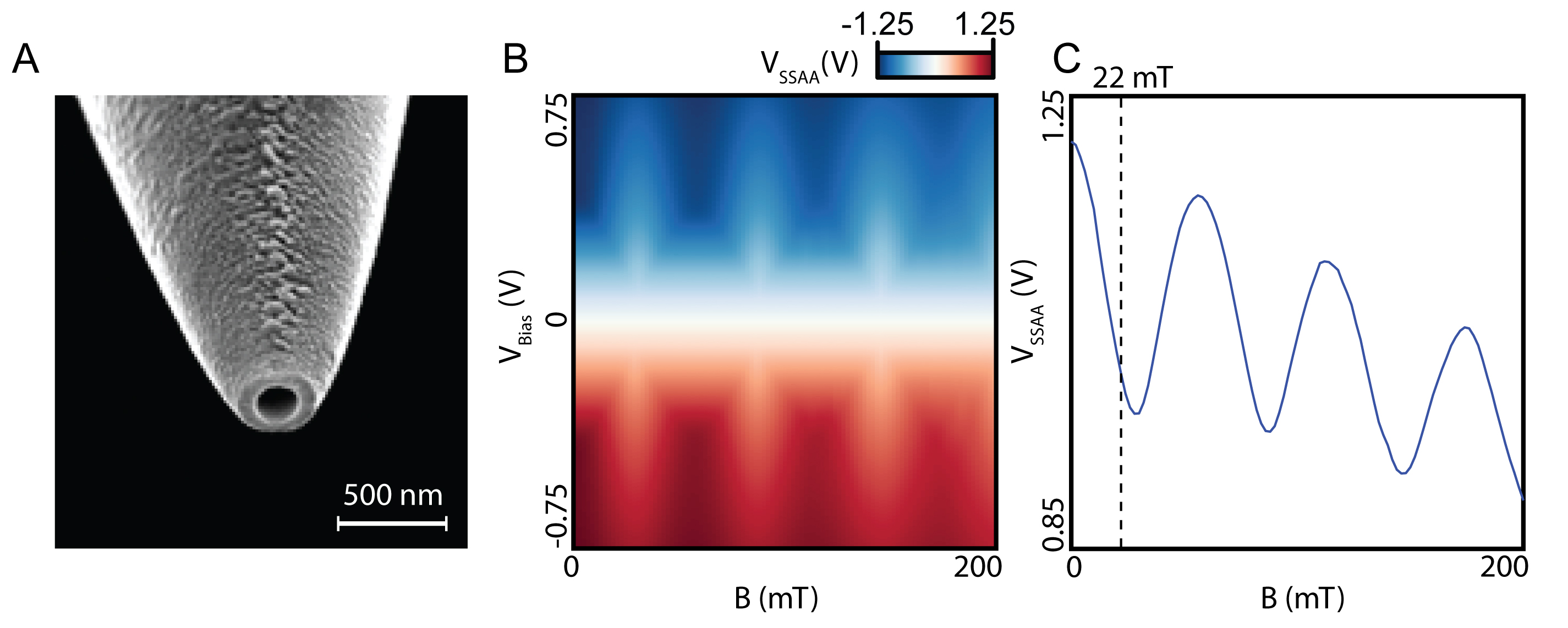}
\caption{
\textbf{NanoSQUID on tip sensors}
(\textbf{A}) Representative SEM micrograph of nanoSQUID sensor. Indium contacts have been thermally evaporated onto each side and onto the tip of the quartz micropipette.
(\textbf{B}) Dependence of electronic properties of nanoSQUID on bias voltage and magnetic field.  Current through the nanoSQUID is amplified by a Series SQUID Array Amplifier (SSAA), the output of which is presented on the color axis.  Indium SQUIDs reliably operate in the 0 - 200 mT range, and usually operate up to 400 mT.  This nanoSQUID was used to generate Figs. \ref{fig:imaging} and \ref{fig:grains}.
(\textbf{C}) Magnetic field dependence of amplified SQUID current at the bias voltage used to generate Figs. \ref{fig:imaging} and \ref{fig:grains}
.}
\label{fig:nanosquid}
\end{figure}

We perform magnetic imaging using nanoscale superconducting quantum interference devices (SQUIDs) fabricated at the tip of a quartz pipette\cite{finkler_self-aligned_2010}.
The SQUID sensors are fabricated by depositing indium on the pulled quartz tubes using a self-aligned three-angle evaporation performed with the tip at cryogenic temperatures\cite{anahory_squid--tip_2020}.
A shunt resistor of $R\approx 3~\Omega$ is integrated near the end of the tip\cite{ vasyukov_scanning_2013}.
A representative scanning electron micrograph of the sensor is shown in Fig. \ref{fig:nanosquid}A.
Indium nanoSQUIDs are maximally sensitive for applied fields below 200 mT for the operating temperature range of the microscope.
The SQUID current is detected using a cryogenic SQUID series array amplifier for optimal noise performance\cite{huber_dc_2001, finkler_scanning_2012, finkler_self-aligned_2010}. At the magnetic fields $B \approx 20 - 50$ mT used in this study, the SQUIDs displayed flux noise down to $250\times10^{-9}$ $\phi_0$ / Hz$^{1/2}$ (where $\phi_0$ is the superconducting flux quantum, $\phi_0 = h/2e = 2 \times 10^{-7}$G cm$^{-2}$) and magnetic field noise down to 15 nT / Hz$^{1/2}$ at frequencies above $\approx 1$ kHz. Fig. \ref{fig:nanosquid}C and D show the low magnetic field interference pattern for the indium SQUID used for the measurements presented in Figs. \ref{fig:imaging} and \ref{fig:grains}.

The magnetic imaging was performed in a pumped helium cryostat at temperatures ranging from 1.6 - 2.2 K. The microscope is housed in a hermetic chamber that is separate from the inner vacuum can, allowing variable pressures of exchange gas to be introduced to the microscopy chamber.
Experiments probing magnetism are performed in high vacuum, $P < 10^{-5}$ mBar, as measuring in larger pressures produces spurious signals related to thermal coupling between the nanoSQUID sensor and sample. During these measurements, the tip is raster scanned at a fixed height above the sample, as determined using the tuning fork based shear force microscopy described in the next section.  For all measurements shown in the main text, scanning was done at a height of 100 nm above the surface of the hBN, for a total height of $140$ nm above the plane of the twisted bilayer graphene.

\subsection{Gradient magnetometry}

\begin{figure*}[ht]
\includegraphics[width=3in]{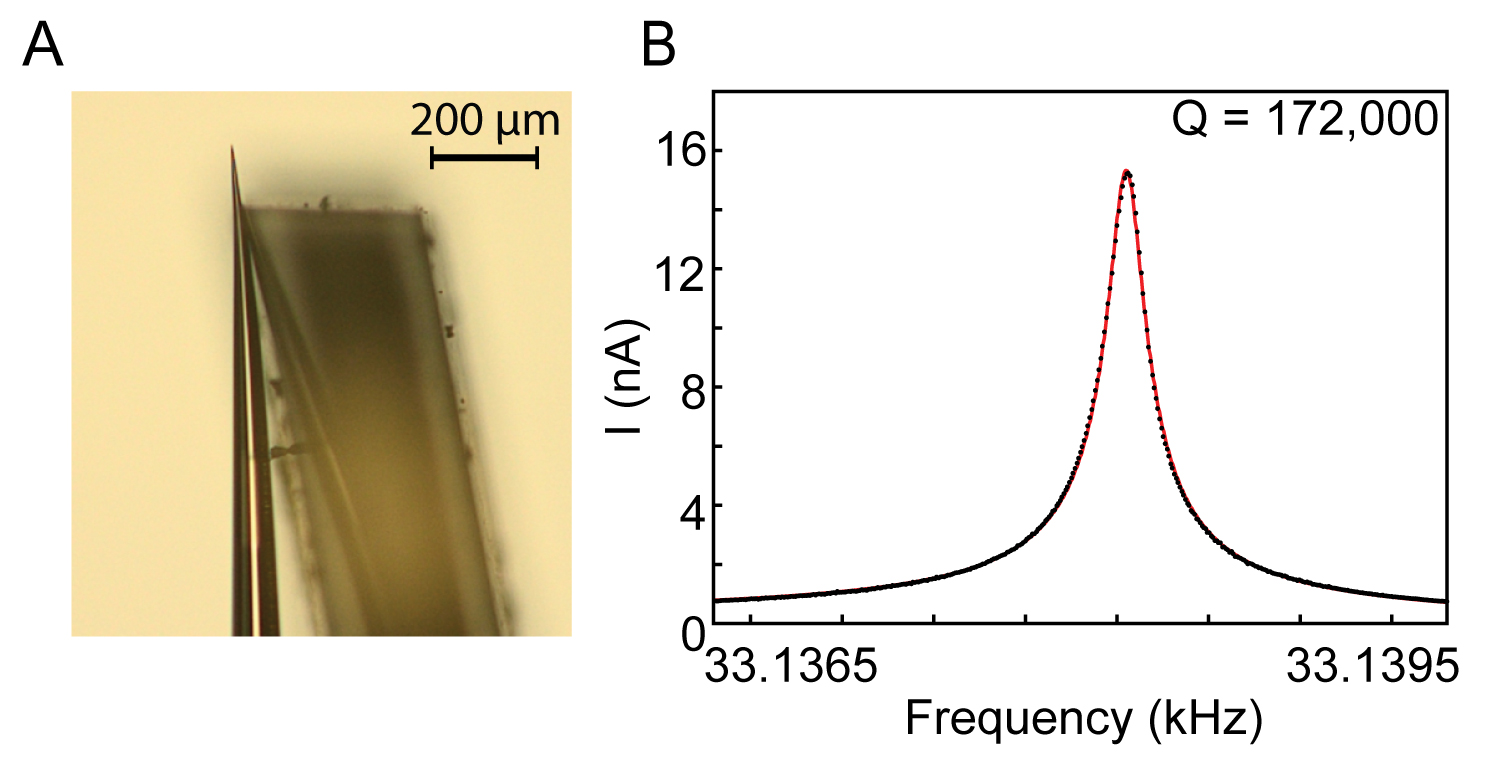}
\caption{
\textbf{NanoSQUID with tuning fork}
(\textbf{A}) Optical image of piezoelectric quartz tuning fork in mechanical contact with a nanoSQUID sensor.
(\textbf{B}) AC current response of tuning fork to $150~\mu\text{V}$ excitation as a function of frequency near resonance. Fit to a Butterworth-Van Dyke model is overlaid.
}
\label{fig:tf_intro}
\end{figure*}

The fringe magnetic fields generated by the tBLG ferromagnet are comparable to the high-frequency noise of our SQUID sensors, but far below the DC noise floor.  As a consequence, it is essential to modulate the input signal at finite frequency.
We contact the nanoSQUID with a piezoelectric quartz tuning fork as shown in Fig. \ref{fig:tf_intro}A\cite{halbertal_imaging_2017}. The tuning fork is electrically excited at the resonant frequency of $\approx33$ kHz, and the AC current through the tuning fork detected using a room-temperature transimpedance amplifier based on Ref. \cite{kleinbaum_note_2012}.  Typical quality factors of the combined SQUID/tuning fork system are $Q\approx 10^5$ (see Fig. \ref{fig:tf_intro}B).
In addition to providing a means of topographic feedback through the tuning fork response, the lateral oscillations of the tip position produce a SQUID response, $B_\tf \approx \hat a\cdot \vec \nabla_r B_z$, that is proportional to the spatial derivative of the static out of plane magnetic field in the direction of the oscillation as described in the main text.

\begin{figure*}[ht]
\includegraphics[width=6.5in]{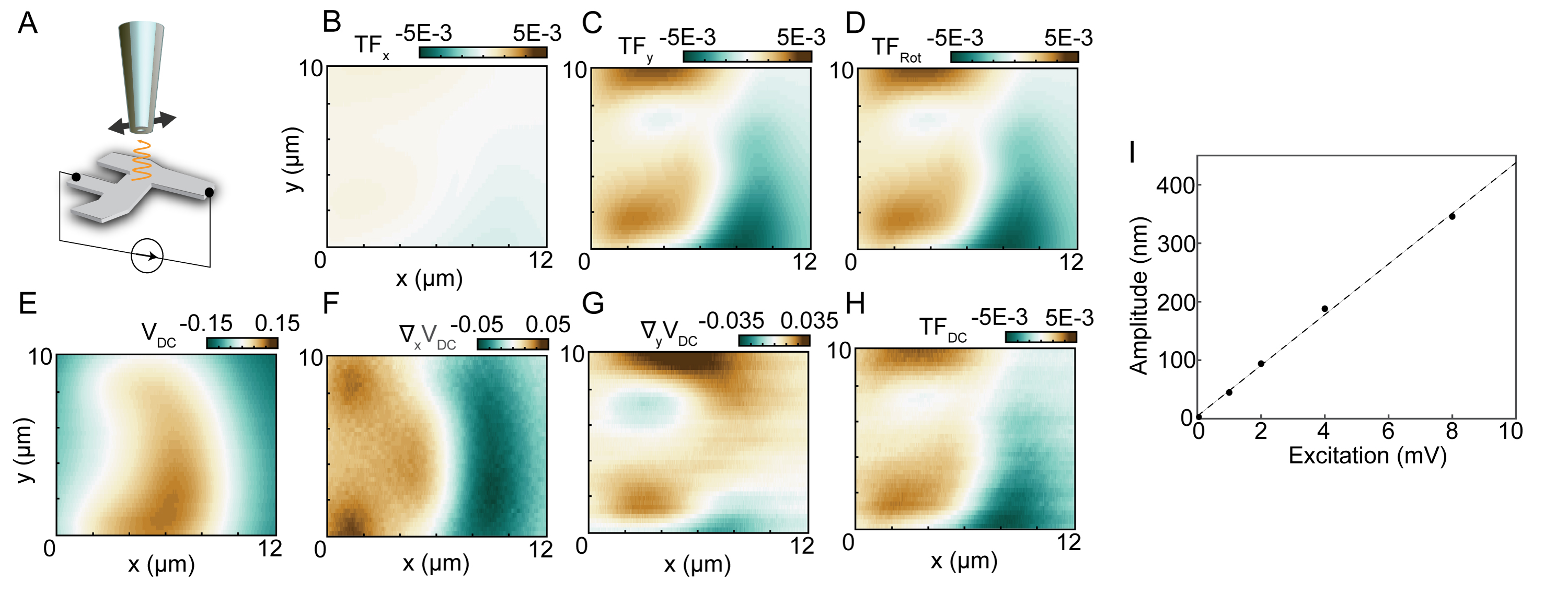}
\caption{
\textbf{AC tuning fork nanoSQUID measurement.}
(\textbf{A}) An applied DC current generates a static temperature distribution above the sample to which the nanoSQUID is sensitive.
(\textbf{B}-\textbf{C})  The sample is scanned under these conditions; the tuning fork AC gradient thermometry signal is shown.  On resonance, the tuning fork gradient signal is primarily in the y quadrature.
(\textbf{D}) A phase rotation is performed on the tuning fork signal to account for small corrections from the x quadrature.
(\textbf{E})  The tuning fork can be calibrated through comparison to a simultaneous measurement of the DC temperature distribution.
(\textbf{F}-\textbf{G})  Gradients of the DC temperature distribution are computed.
(\textbf{H})  A least squares fit of the AC gradient thermometry signal in \textbf{D} to a linear combination of the gradients of the DC temperature distribution \textbf{E} and \textbf{F}.  These scans correspond to a tuning fork amplitude of $346\pm7$nm and an angle relative to the y axis of $19.9\pm0.2^\circ$.
(\textbf{I}) Amplitude of oscillation of SQUID as a function of tuning fork drive voltage with a linear fit.
}
\label{fig:tf_cal}
\end{figure*}

Quantitative measurements using $B_{TF}$ require having a proper calibration of $\hat a$, the vector describing the nanoSQUID oscillation.
This can be measured by fitting a DC signal to the simultaneously measured AC gradient signal. Naturally, this requires a DC signal sufficiently large that it can be easily measured by the nanoSQUID. Given that there are no sufficiently strong sources of magnetic field close to the tBLG heterostructure, we elected to do the calibration by taking advantage of the nanoSQUID's thermal sensitivity\cite{halbertal_imaging_2017}. The microscopy chamber is flooded with $P\approx 0.1 - 1~ \text{mBar}$ of helium gas to facilitate heat exchange between the nanoSQUID and sample while a DC current is applied through the graphene heterostructure (Fig. \ref{fig:tf_cal}A). The simultaneously acquired in- and out-of-phase AC signals at the resonant frequency of the tuning fork are shown in Figs. \ref{fig:tf_cal}B and \ref{fig:tf_cal}C.  The nanoSQUID signal at the tuning fork frequency is in phase with the position of the tuning fork.  Like any damped harmonic oscillator, close to resonance the position is out-of-phase with the driving force. We therefore expect to find the nanoSQUID signal at the tuning fork frequency to be out-of-phase with the drive voltage, and indeed that is what is observed. To avoid errors stemming from small detuning of our drive frequency from the resonant frequency, we perform a phase rotation to capture the signal from both quadratures (Fig. \ref{fig:tf_cal}D).

Assuming the tuning fork amplitude is small relative to length scales of thermal gradients, the SQUID response at the tuning fork frequency obeys:
\begin{equation}
    V_{TF} = \hat a \cdot \vec \nabla_r V_{DC}
\end{equation}
From the measured DC signal (Fig. \ref{fig:tf_cal}E), we numerically calculate $\nabla_{x}V_{DC}$ and $\nabla_{y}V_{DC}$ (Figs. \ref{fig:tf_cal}F and \ref{fig:tf_cal}G).  A least-squares fit over $\hat{a}$ to $V_{TF}$ according to equation S1 produces an excellent fit (Fig. \ref{fig:tf_cal}H). This process is repeated for several amplitudes of the tuning fork drive voltage.  We find that the displacement amplitude is linear in the applied voltage, as expected (Fig. \ref{fig:tf_cal}I).

The oscillation amplitude calibrations are performed in rarefied helium gas, whereas the magnetic measurements are performed in high vacuum. Changes in ambient pressure affect the damping in the tuning fork and can therefore change the amplitude for a given excitation voltage. This effect is captured by the quality factor of the resonance, which is expected to increase with decreasing ambient pressure. The oscillation amplitude is linear in both the excitation amplitude and the quality factor Q near resonance when in the linear driving regime\cite{dagata_scanning_1992}. During the AC gradient thermometry measurements presented in Fig. \ref{fig:tf_intro}, the quality factor of the tuning fork-nanoSQUID system was
$Q_{gas} = 89\times 10^3$.
Immediately after the heat exchange gas was removed, the quality factor increased to $Q_{min} = 120\times 10^3$.  Over the course of the several week experiment during which the microscopy chamber was continuously pumped, the quality factor of the tuning fork-nanoSQUID assembly slowly increased, ultimately reaching $Q_{max} = 172\times 10^3$. Every AC gradient magnetometry data set presented here is assumed to have the measured calibration modulated by a factor falling in the range $|\hat{a}_{vac}| \in (\frac{Q_{min}}{Q_{gas}},\frac{Q_{max}}{Q_{gas}})\cdot |\hat{a}_{gas}| $, or $|\hat{a}_{vac}| \in (1.35,1.93)\cdot |\hat{a}_{gas}|$. This systematic uncertainty dominates other sources of error, and produces the upper and lower bounds on $m$ presented in Fig. \ref{fig:imaging}H.  We note that this error could be made negligible in future experiments by more frequent calibrations of the tuning fork $Q$.

%Outline:
%1. It's important to do these measurements at high frequency (1/f noise, small magnetization)
%2. Normal AC modulation doesn't work well because (a) these states are metastable (b) preparing them involves large excursions in B, and B can't be substantially varied rapidly.
%3. NanoSQUID is way too big to be mounted on a tuning fork -> shear AFM.
%4. Discuss uncertainty in TF oscillation amplitude

\subsection{Reconstructing magnetization from measured $B_{TF}$}

The SQUID-tuning fork measurement, for oscillation amplitudes smaller than magnetic features, gives $B_\tf \approx \hat a\cdot \vec \nabla_r B_z$. Once we have determined $\hat a$, as explained in the previous section, we can recover $B_{z}$ and consequently the magnetization density. All subsequent operations are performed on difference datasets to ensure that $B_{TF}$ is purely magnetic, and doesn't have residual contributions from temperature or electric field.

To integrate the data, we take the bottom and left edges of each dataset to have the boundary condition of zero magnetic field. For each datapoint in a two dimensional map of $B_{TF}$, we interpolate a line from the bottom or left edge along the direction of $\hat a$ with cubic interpolation taking approximately one data point per pixel crossed. We then numerically integrate that line using trapezoidal integration. This process is repeated for every datapoint in the dataset.

 The resulting dataset is a two dimensional magnetic field map $\mathbf{B_{z}}(x,y,z=z_0)$ at height $z_0$. In order to calculate the magnetization density of the sample $\mathbf{m}(x,y,z=0)$, we performed a numerical reverse propagation of the Fourier transform of the magnetic field, based on  \cite{thiel_probing_2019,roth_using_1989}. From Maxwell's equations,
\begin{equation}
    \nabla \times \mathbf{H}=0,
\end{equation}
and
\begin{equation}
    \nabla \cdot \mathbf{H} = -\nabla \cdot \mathbf{m}
\end{equation}
Outside the sample, $\nabla \cdot \mathbf{H} = 0$ and we define the potential function $\phi_m(\mathbf{k},z)$ in Fourier space where $\mathbf{H} = -\nabla\ \phi_m$, and $\mathbf{k} = (k_x, k_y)$, where $k_x$ and $k_y$ are the corresponding coordinates of $x$ and $y$ in Fourier space.

By solving Maxwell's equations in Fourier space, $\phi_m(\mathbf{k},z)$ is given by

\begin{equation}
    \phi_m(\mathbf{k},z) = \frac{\sigma(\mathbf{k},0)}{2}e^{-kz},
\end{equation}
and therefore the perpendicular component of the magnetic field $ {B}_{z}(\mathbf{k},z)$ is given by

\begin{equation}
    {B}_{z}(\mathbf{k},z) = \mu_0 k\frac{e^{-kz}}{2}\sigma(\mathbf{k},0),
\end{equation}
where $\sigma(\mathbf{k},0)$ is the surface moment density of the sample.
%here I did not mention perpendicular component and other components of the field%
The DC signal detected by the nanoSQUID, $B_\mathrm{z}$, is the out-of-plane magnetic field averaged over the area of the nanoSQUID. In Fourier space, this convolution process is equivalent to multiplying a weighting function

\begin{equation}
    h(k) = \frac{J_1(k R)}{kR/2},
\end{equation}
where $J_1$ is the first order Bessel function of the first kind and $R$ is the effective radius of the nanoSQUID sensor, extracted from the periodicity of the nanoSQUID response in an applied magnetic field (see Fig. \ref{fig:nanosquid}C). Therefore, the SQUID signal is given by

\begin{equation}
    B_{\mathrm{z}}(\mathbf{k},z) = h(k)\mu_0 k\frac{e^{-kz}}{2}\sigma(\mathbf{k},0),
\end{equation}

The averaging effect of the nanoSQUID diameter is convoluted with the averaging effect of the height of the nanoSQUID $h$ above the tBLG sample.  Together these length scales limit the cutoff frequency of the magnetic signal that our nanoSQUID sensors can resolve, $k_{\max} = \min(2\pi/h, 3.83/R)$\cite{thiel_probing_2019,roth_using_1989}. In order to comply with this limit and filter out the high frequency noise, we apply a Hanning window to the SQUID signal,

\begin{equation}
W(k)=\left\{
        \begin{array}{ll}
            \frac{1}{2}\left(1+\cos\frac{kh}{2}\right),   &  k<k_{max}\\
             0,                                              &  k\geq k_{max}.
        \end{array}
    \right.
\end{equation}

From equation (S5), we obtain the surface moment density of our sample,

\begin{equation}
    \sigma(\mathbf{k},0) = \frac{W(k)\cdot B_{\mathrm{z}}(\mathbf{k},h)}{\mu_0 k\cdot(e^{-kz}/2)}.
\end{equation}

\begin{figure*}[ht]
 \includegraphics[width=4.5in]{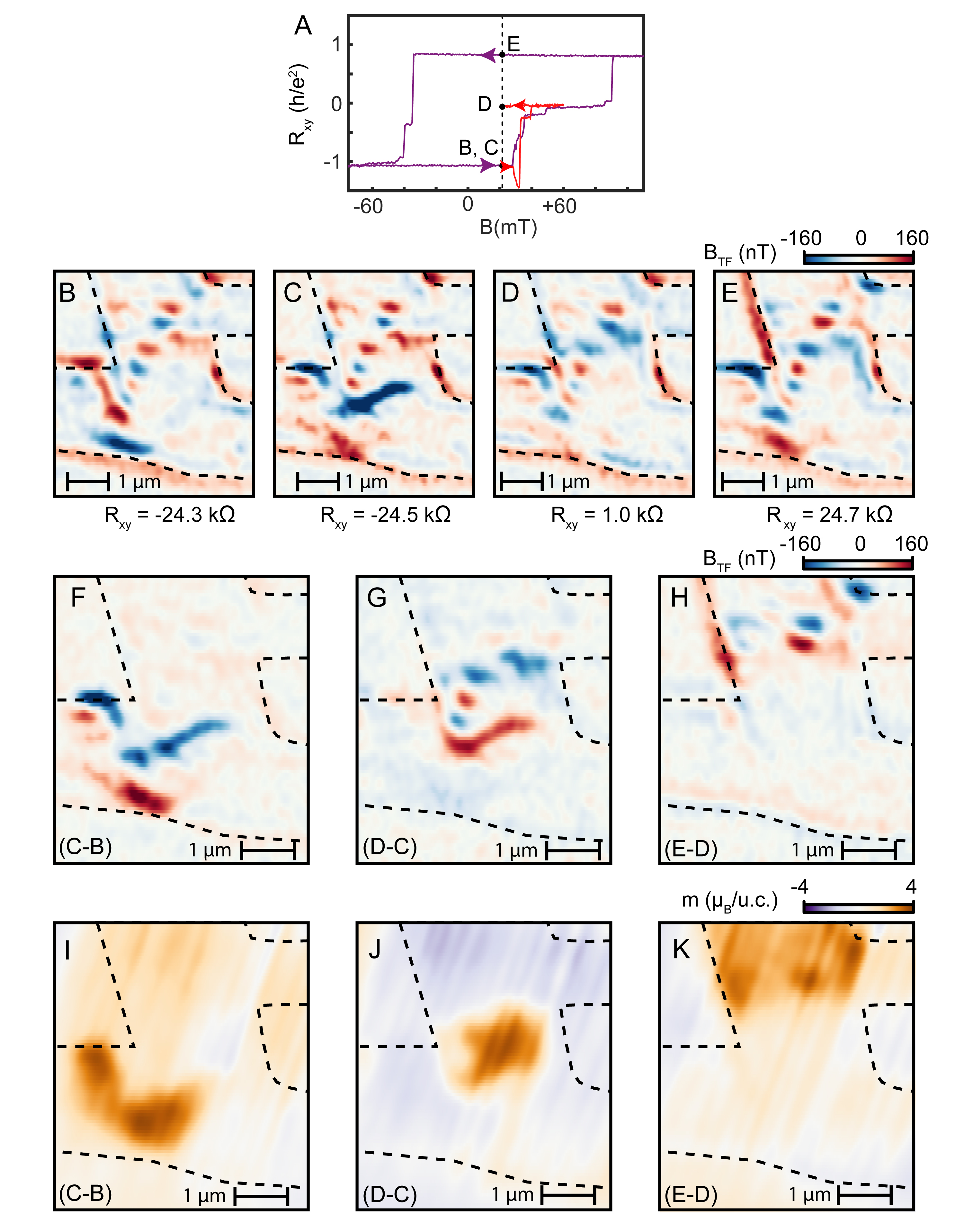}
\caption{
\textbf{Additional images of  the magnetic domain wall motion.}
(\textbf{A}) Hall resistance $R_{xy}$, showing the preparation of magnetic domain configuration imaged in (\textbf{B-E}).
(\textbf{F-H}) show magnetic images obtained by taking a pairwise difference of images (\textbf{B-E}), as indicated in the bottom-left corners. These images demonstrate the motion of  domain walls.
(\textbf{I-K}) Reconstructed magnetization density corresponding to (\textbf{F-H}).  Inversion of the domain shown in \textbf{I} does not have a substantial impact on $R_{xy}$ because the domain does not separate any pairs of contacts.
}
\label{fig:S:moregrains}
\end{figure*}

This is plotted as the magnetization $m$ in Figs. \ref{fig:imaging}, \ref{fig:grains}, and \ref{fig:S:moregrains}.

\subsection{Parasitic contributions to $B_\tf$}

In the main text, we see that unprocessed $B_{TF}$ scans have a ubiquitous non-hysteretic signal at the edge of the device. In this section, we provide evidence that this signal is neither magnetic in origin, nor related to the physics of twisted bilayer graphene.

\begin{figure*}[ht]
\includegraphics[width=2.75in]{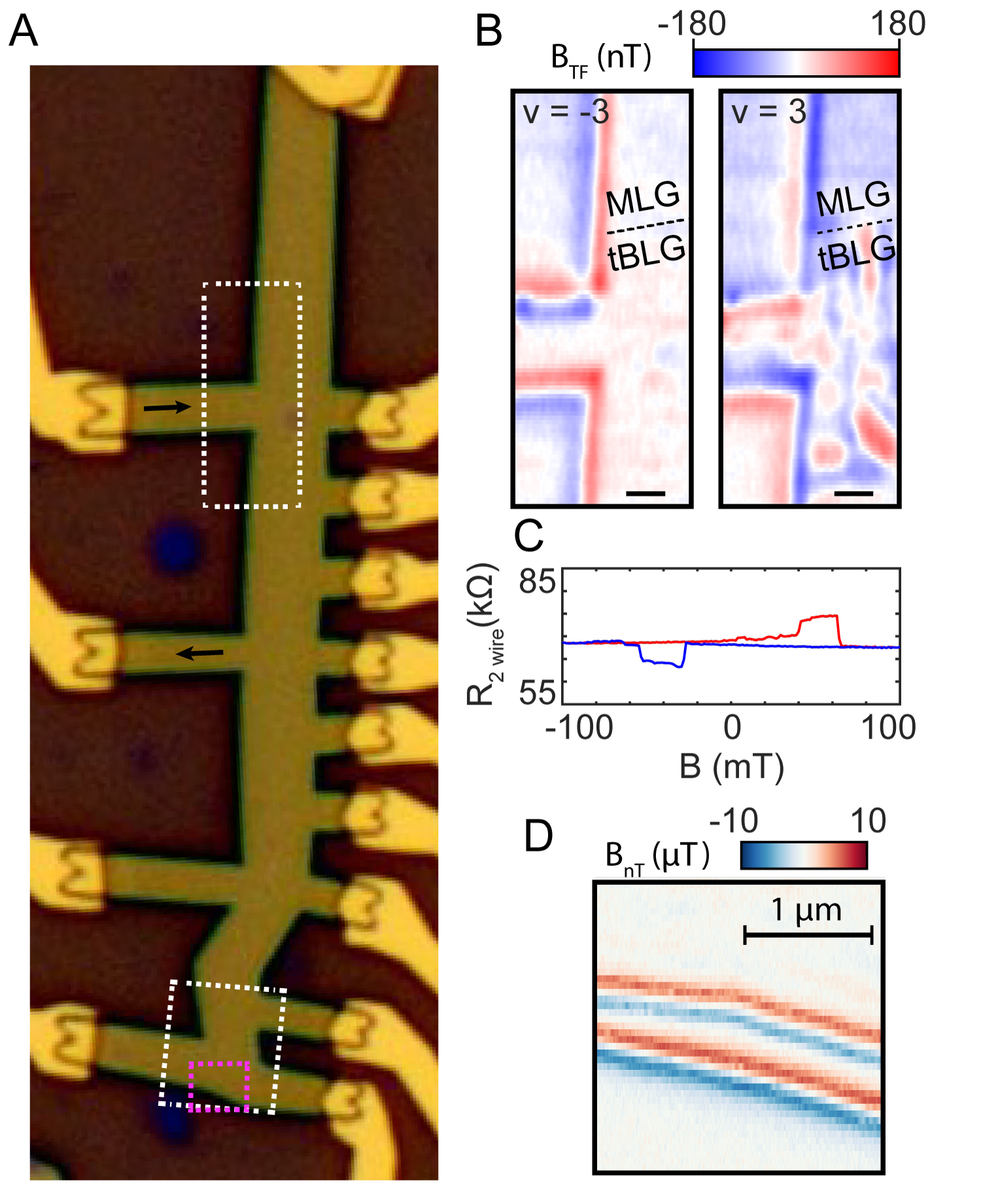}
\caption{
\textbf{Parasitic electric field sensitivity of nanoSQUID sensor.}
(\textbf{A}) Optical image of entire tBLG device.
(\textbf{B})  NanoSQUID AC gradient magnetometry of region of device far from QAH region. Scans at filling factors $\nu = -3$ electrons per unit cell and $\nu = 3$ electrons per unit cell are shown.  Disordered magnetic structure is visible in the bulk of this region of the device at $\nu = 3$ but not at $\nu = -3$.  Magnetic order ends at the edge of one of the monolayers used to produce the tBLG. In both images a signal comparable in size to the magnetic signal is observed at all edges of the device, where electric fields from the gate are strongest.  This signal changes sign when the applied gate voltage changes sign. Scale bar is $1~\mu m$.
(\textbf{C}) Ferromagnetism observable in 2 wire transport measurement through region imaged in B at $\nu = 3$.  Source and drain contacts are shown with black arrows overlaid on A.
(\textbf{D}) Image of $B_\tf$ in region marked with maroon square overlaid on A.  The nanoSQUID sensor used for this scan had unusually pronounced contrast at the edge of the device. We report this scan in units of T for comparison to \textbf{B}, though we do not believe that the signal is magnetic in origin.
}
\label{fig:parasitics}
\end{figure*}

Fig. \ref{fig:parasitics}A shows an optical image of the full device. The bottom white square highlights the region that exhibits the quantum anomalous Hall effect; this is the region in which all the scans presented in the main text are performed. The white rectangular overlay at the top of the device encompasses a region in which there is an edge of one of the graphene layers forming the tBLG. The top part of this region is gated monolayer graphene (MLG), whereas the bottom is gated tBLG.
Fig. \ref{fig:parasitics}B shows images of $B_{TF}$ taken in the top region when gated to both $\nu = +3$ and $\nu = -3$. Magnetism is present only at $\nu = +3$, and only in the tBLG region. Fig. \ref{fig:parasitics}C shows two terminal transport hysteresis curves taken from the contacts indicated in Fig. \ref{fig:parasitics}A, consistent with the scanning observation of disordered magnetism. The edge signal is still present along the edge of the MLG part of the device, thus confirming that it is not associated with the topological band structure of the tBLG. Furthermore, the sign of the edge signal flips when gated to the opposite sign. This is more simply explained by a sensitivity to local electric field than by exotic edge magnetism with gate dependent field coupling.

We have observed that nanoSQUID sensors differ in the strength of their coupling to the contrast mechanism present at the edges of gated heterostructures. The majority have similar qualitative behaviors as those presented in the main text and Figure \ref{fig:parasitics}B. Occasionally, however, some nanoSQUID sensors exhibit different and much stronger responses to edges of gated heterostructures. Figure \ref{fig:parasitics}D is a scan of the region marked with the maroon square with such a nanoSQUID sensor. Not only is the signal larger relative to the magnetic sensitivity, but we also see a more sophisticated pattern associated with the edge.

%We interpret this to likely be a resonant effect driven by stray electric fields at the edge of the device.

Though the exact mechanism is unknown, one possible explanation is that this contrast mechanism corresponds to electric field modulation of the properties of the nanoSQUID sensor. We cannot, however, rule out a thermal component to the contrast mechanism. Nevertheless, as is explained above, the evidence is strong that this signal arising at the edge is non-magnetic in origin and is not related  to the  physics of orbital ferromagnets and edge states that we report in this paper.

\clearpage

\section{Supplementary Data}

\begin{figure*}[ht]
\includegraphics[width=7.25in]{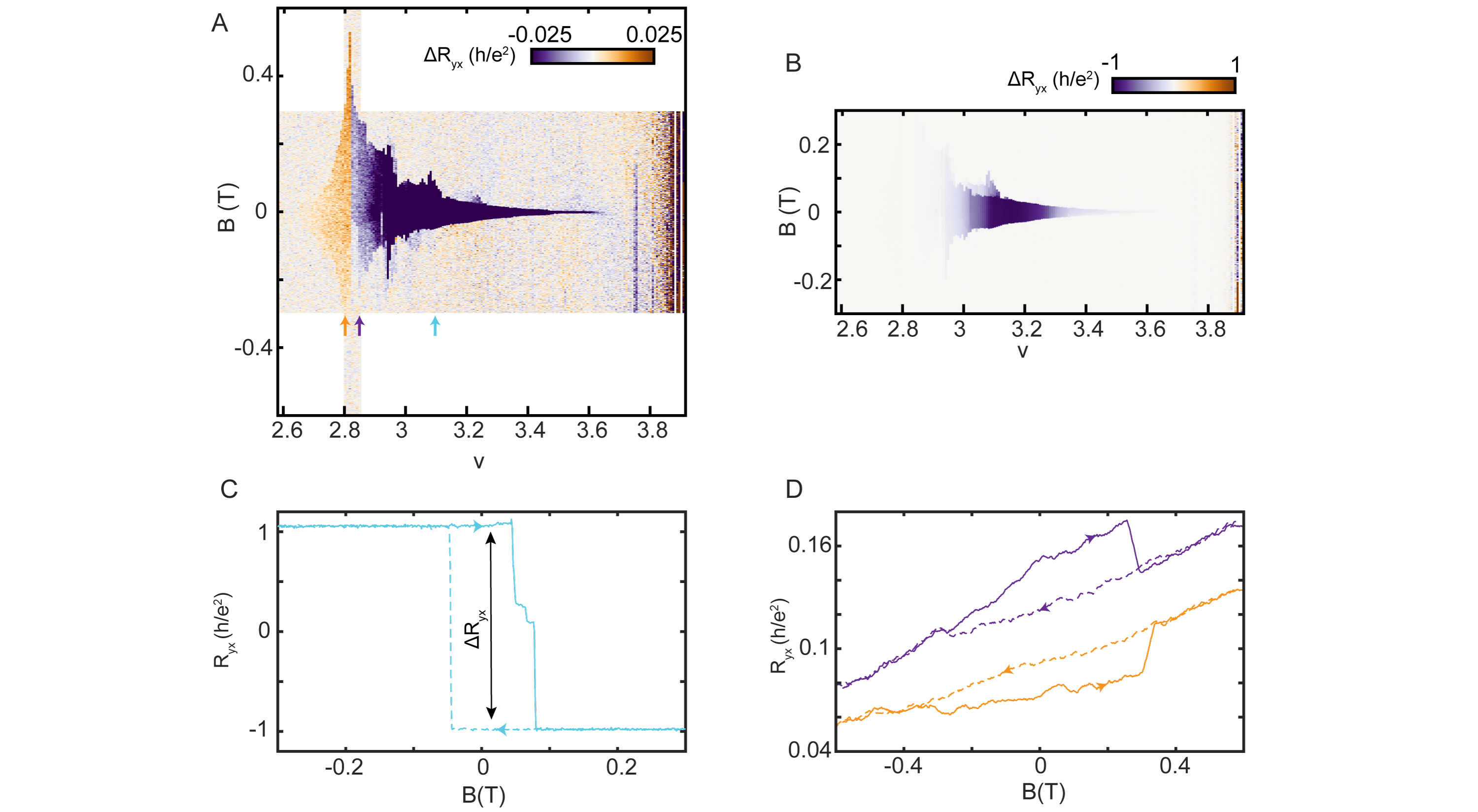}
\caption{
\textbf{Valley polarization switching in tBLG.}
(\textbf{A}) Anomalous Hall resistance $\Delta R_{yx}$ associated with twisted bilayer graphene ferromagnetism, extracted by subtracting $R_{yx}(B)$ as $B$ is increased from $R_{yx}(B)$ as $B$ is decreased.  The colorscale is chosen to illustrate weak features in $\Delta R_{yx}(\nu)$.  For $\nu < 3$, the coercive field of the ferromagnetic order increases dramatically, peaking at $\nu = 2.82$ electrons per moir\'e unit cell. For $\nu < 2.82$, $\Delta R_{yx}$ switches sign, indicating that the valley polarization of the ground state of the system at finite magnetic field has switched. The coercive field plotted in Fig. \ref{fig:density}C is extracted from this dataset.  (\textbf{B}) shows the same data as (A), but with the colorscale fixed to the von Klitzing constant to show the range of filling factors for which a robust QAH effect is observed.
(\textbf{C})  Robust Chern 1 QAH effect at $\nu = 3.1$.
(\textbf{D}) Ferromagnetic hysteresis plots on opposite sides of the divergence of the coercive field close to $\nu = 2.82$ (with offset).  Note the change in the relative sign of $\Delta R_{yx}$. }
\label{fig:S:signchange}
\end{figure*}

\begin{figure*}[ht]
\includegraphics[width=4.5in]{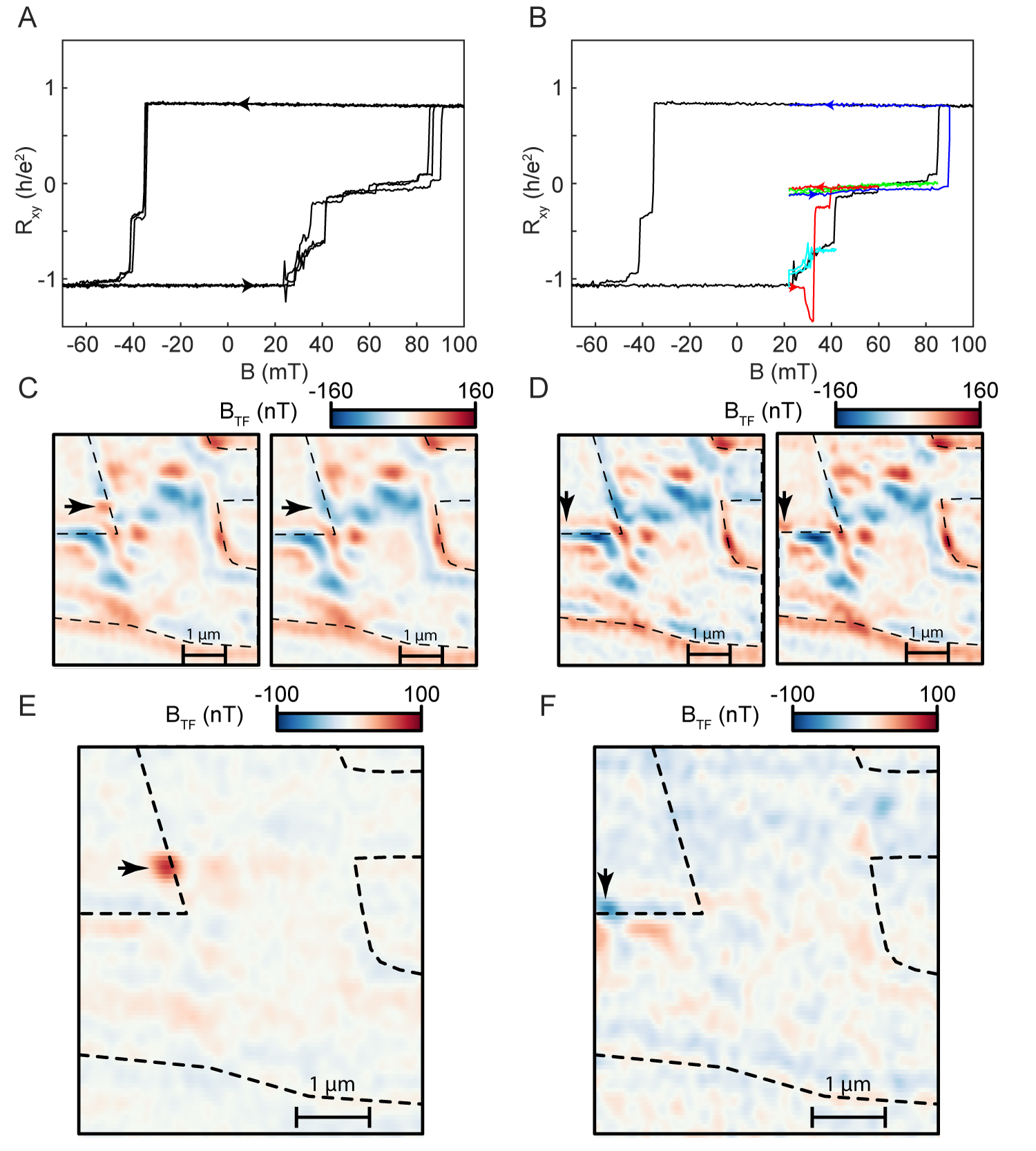}
\caption{
\textbf{Subresolution domain reversals in the $R_{xy}\approx 0$ plateau}
(\textbf{A}) $R_{xy}$ is repeatedly measured through a complete major loop of the tBLG ferromagnet. Subtle differences appear between each of the three hysteresis loops, but all share the $R_{xy}\approx0$ plateau.
(\textbf{B}) $R_{xy}$ is measured through a set of minor loops of the ferromagnetic hysteresis curve.  The cyan loop occurs over a regime in which changes in $R_{xy}$ as a function of magnetic field are reversible.  The red loop occurs over the transition from the fully polarized field-unfavored state to the mixed state with $R_{xy}\approx0$ discussed in Fig. \ref{fig:grains}.  The $R_{xy}\approx0$ plateau contains several states characterized by small but measurable changes in $R_{xy}$.  The green loop characterizes one such small change in $R_{xy}$.  The blue loop characterizes the transiton from the mixed state with $R_{xy}\approx0$ to the fully polarized field-favored state.
(\textbf{C}) Scans were performed before and after the green minor loop in \textbf{B}, and are shown here.  The magnetic field distributions are qualitatively similar with a few small differences.
(\textbf{D}) An additional pair of scans with similar but unequal $R_{xy}$ prepared within the $R_{xy}\approx0$ plateau.
(\textbf{E-F}) Difference images corresponding to \textbf{C-D}.  All states within the $R_{xy}\approx0$ plateau had domain distributions qualitatively similar to that discussed in Fig. \ref{fig:grains}K.  Changes in the precise value of $R_{xy}$ correspond to sub-resolution changes in the magnetic field distribution above the tBLG device.
}
\label{fig:trans_repeat}
\end{figure*}

\clearpage

\end{document}